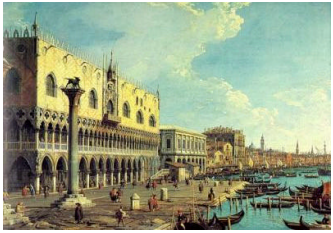
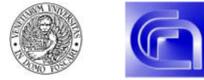

# Euro PVM MPI 2003

Venezia, San Servolo,
Sep. 29 - Oct. 2

## Message Passing Fluids: molecules as processes in parallel computational fluids

Gianluca Argentini

*New Technologies & Models*
Information & Communication Technology Department
*Riello Group*, Legnago (Verona), Italy

**1  Message Passing Fluid**

- an abstract computational model on fluids
- molecules move on the basis of some predefined rules on a 2D or 3D grid
- single molecules receive physical information from other near molecules
- information exchanged is necessary for the computation of the movement to the next cell

**2  Physical information exchanged**

- positions of molecules in the grid
- physical quantities (e.g. Van der Waals interaction)
- presence of obstacles (e.g. external walls)

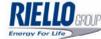

= a molecule with its euclidean neighbourhood
= some molecules of fluid
= cells occupied by an external wall

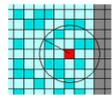

*2D example of physical neighbourhood for a molecule in a MPF*

**3  Possible rules for the movement**

- only one molecule on one single cell (**1m1c**)
- a molecule moves to the cell of its *geometric neighbourhood* where the sum of physical potentials due to other objects of its *physical neighbourhood* is minimum
- other molecules and walls act with repulsive forces on a single molecule
- compliance with physical conservation laws

*example of geometric 2D Moore 8-neighbourhood for a molecule*

**1, 2, 3** ➡ an MPF is essentially a *generalized Cellular Automaton*

**4  a.  A parallel computational model**

- $P$ = number of parallel processes (processors)
- $N$ = number of molecules, such that $N = M\,P$ with $M$ integer
- every process receives at the beginning its $M$ molecules
- at every computational step a process sends the physical information about its molecules to the *communicator group* and receives the same kind of messages from other processes
- for each molecule the process, on the basis of predefined rules, computes the physical parameters affected by the molecules of the neighbourhood
- the grid is updated with the new positions of all molecules

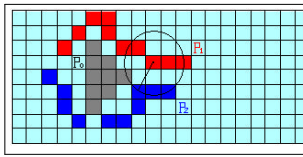

$p_0$ = process of grey molecules (walls)
$p_1$ = process of red molecules
$p_2$ = process of blue molecules

*the centered molecule has 1 grey, 4 red and 3 blue molecules on its neighbourhood*

**4  b.  Using MPI**

- every cell in the grid is identified by a single integer
- use of a two-dimensional array as grid variable for storing the positions of the $M = N/P$ molecules of each process:

```
int global_positions[P][M];
```

- use of a one-dimensional array as local variable of a single process for storing the positions of its $M$ molecules:

```
int local_positions[M];
```

ex.: local_positions[2] = 50 assigns the *3rd* molecule of the current process to position 50 in the grid

- at the beginning the *rank 0* process distributes the molecules among all other processes:

```
MPI_Scatter(&global_positions[0][0],M,MPI_INT,local_positions,M,MPI_INT,
0,MPI_COMM_WORLD);
```

- at every step all processes update the grid variable of the molecules after the computation of the new positions:

```
MPI_Allgather(local_positions,M,MPI_INT,global_positions,M,MPI_INT,
MPI_COMM_WORLD);
```

- check of condition **1m1c**: `MPI_Send` from *rank 0* process to others (**RMA** not tested)

**4  c.  Some critical aspect of the model**

- the computational effort of every process for determining the neighbourhoods of its molecules and performing the calculations necessary for the new positions can be high for large values of $N$ (computational complexity order is $\sim (k_1 T_{calc} N + k_2 T_{comm} N^2)$
- the use of a grid variable updated by a call to *MPI_Allgather* can be expensive in terms of memory usage
- the effort for the compliance of the condition **1m1c** can be expensive if the *density* of the molecules in the grid is high

### First experiments

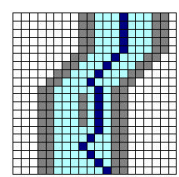

*Trajectory of one molecule (blue) in a channel with internal obstacle (grey)*

*Rule: at every step the molecule moves to the cell where the newtonian potential due to other objects is minimum*

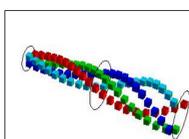

*Trajectories in a turbine simulation*

*Rules: at every step a molecule moves in the 3D direction determined by a sinusoidal force and in that cell of the normal plane where a newtonian potential is minimum*

**condition 1m1c seems satisfied better with use of some kind of shared memory**

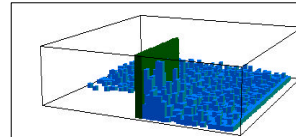

*Cluster of molecules (blue) in a box with internal obstacle (green)*

*Rule: at every step a molecule moves to the cell where the newtonian potential due to other molecules and objects is minimum; if a cell has more than one particle, the molecules are vertically stacked*

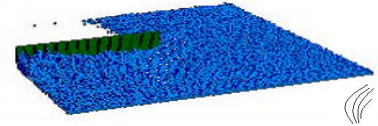

*The same simulation with a larger N: the blank dots and the black lines in the corner show the curvature of the geometric disposition of molecules induced by the obstacle*

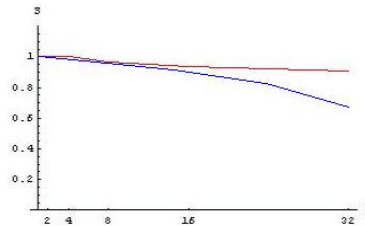

*P (= processors) - S (= efficiency) diagram from experiments for this computational model*

*Hardware and software: cluster Linux, Intel 1.33 GHz processors, GNU C-compiler 2.96, MPICH 1.2.4 libraries (no shmem) [ CINECA, Bologna, Italy ]*

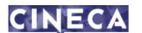

**red line** : without check of **1m1c** condition
**blue line** : with check of **1m1c** condition

**5  MPF application in Computational Fluid Dynamics**

- *Classical CFD*: geometrization of the problem, mesh or grid construction, discretization of Navier-Stokes PDEs, computation of solution
- *Non standard CFD*: grid, geometry and movement rules definition, computation of trajectories by adaptive algorithm, physical interpretation of results
  - wide use of Cellular Automata methods in CFD from 1985
  - flexibility of rules
  - the solution grows in the time adapting themself to the geometry of the problem
  - MPF model has computational steps easily tractable by parallel libraries and architettures (SMP or multinodes hardware)
  - MPF model has a communication algorithm which can be tested by message passing paradigm or shared memory technique

**Scheme of a program with low communication:**

1. Evolution of the Cellular Automaton in the computational grid (message passing expensive, necessity of an efficient parallel algorithm, computational complexity $\sim (k_1 T_{calc} N + k_2 T_{comm} N)$, $N$ = molecules number, $k_1$ and $k_2$ integers)

2. Computation of smooth trajectories with fine sampling in time (flops computational step, SIMD coding, data distribution on the processors, no message passing, computational complexity $\sim NS^2$, $S$ = sampling dimension for smoothing)

3. Rendering of the graphics (memory expensive, but parallelizable)

**6  a.  Doing CFD with MPF**

- 3D grid
- definition of initial conditions (positions, flux velocity) and boundary conditions
- definition of rules (**1m1c**, geometrical type of neighborhood, direction of motion in presence of obstacles)
- grid variable for tracing the molecules in the grid (0 = free cell, 1 = particle or obstacle)
- possible values for the components of the velocity vector of a molecule: 1, 0, -1

$$x(t_{n+1}) = x(t_n) + p_x$$
$$y(t_{n+1}) = y(t_n) + p_y$$
$$z(t_{n+1}) = z(t_n) + p_z$$

*Algebraic rules for computing the new position of a molecule*

$p_i$ = parameters which depend on the physical status of the fluid and on the predefined rules; in the discrete model the possible values are 1, 0, -1; a possible rule is: at every step the default values are $p_i$ = velocity$_i(t_n)$

**Adaptive algorithm**

*Loop until a molecule moves to a new position or a max predefined value of cycles is reached ( i = x, y, z ):*

*if $u_i(t+1) = u_i(t) + V_i(t)$ is not free, then $V_i(t)$ is changed according to predefined rules*

**6  b.  Some aspect of the computational model**

- for every molecule the program constructs the array of the traced positions
- for more realistic visualization of the results, the points of a trajectory are used as basis for a Bezier or other type of interpolating curves
- for large $N$, for large number of interpolation points and for fine sampling (graphics resolution) the computational effort to obtain the values of the Bezier curves can be very strong

**6  c.  Using MPI**

- multiprocess distribution of molecules based on initial geometric positions
- at every step *rank 0* process checks the condition 1m1c:

`MPI_Barrier`, after positions computation

`MPI_Gather`, for *rank 0* process

`MPI_Send`, from *0* to others if necessary

- at the end of all the computations *rank 0* process distributes (`MPI_Scatter`) to the other processes the Bezier functions for their numeric evaluation on fine sampling

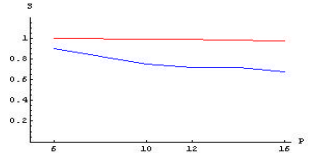

*P (= processors) - S (= efficiency) diagram:*

*Hardware and software: Unix nodes, PA-RISC and SPARC 400 MHz processors, 100 Mb LAN, GNU C-compiler 2.96, MPICH 1.2.4 libraries [ ICT, Riello ]; cluster Linux [ CINECA ]*

max 100 used molecules for every processor

**blue line** : MPI for positions computation (message passing for **1m1c** condition)

**red line** : MPI for Bezier curves evaluation (no message passing, only SIMD step of the algorithm, a sequential Horner's rule performed by a single process)

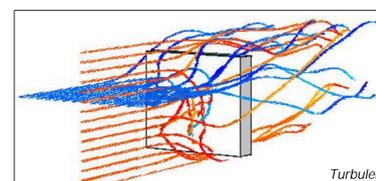

*Some trajectories of a vertical plane and of an horizontal plane of a flow in presence of an obstacle normal to initial velocity (100x40x40 grid, 1600 particles)*

*Turbulence due to mixture of two fluids from opposite motion directions*

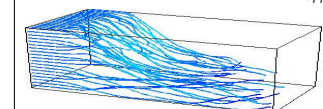

*Trajectories for x-z plane and x-y plane for a flow constrained by a narrowing (80x40x40 grid, 1600 particles)*

**7  Open questions**

- relation between MPF adaptive algorithms and physical *conservation laws*; study and simulation of *turbulence* with MPF
- efficient *message passing* or *shared memory* technique for a good speedup of the positions computation
- comparison between *distributed Horner* and *parallel Horner* for polynomials evaluation
- comparison of *results* and *performances* between parallel Navier-Stokes methods and parallel MPF algebraic scheme in some geometrical and physical situation

Gianluca Argentini, *Riello Group*, via Alpini 1, 37045 Legnago (Verona), Italy - gianluca.argentini@riellogroup.com    **Thanks**